\documentclass[aps,showpacs,prl,twocolumn]{revtex4}
\usepackage{graphicx,epsfig,bm}

\begin{document}

\title{Quantum Speed Limit for Perfect State Transfer in One Dimension}

\date{\today}

\author{Man-Hong Yung}

\email[Electronic address: ]{myung2@uiuc.edu}

\affiliation{Department of Physics, University of Illinois at
Urbana-Champaign, Urbana IL 61801-3080, USA}


\begin{abstract}
The basic idea of spin chain engineering for perfect quantum state
transfer (QST) is to find a set of coupling constants in the
Hamiltonian, such that a particular state initially encoded on one
site will evolve freely to the opposite site without any dynamical
controls. The minimal possible evolution time represents a speed
limit for QST. We prove that the optimal solution is the one
simulating the precession of a spin in a static magnetic field. We
also argue that, at least for solid-state systems where interactions
are local, it is more realistic to characterize the computation
power by the couplings than the initial energy.

\end{abstract}

\maketitle


The transfer of a quantum state from one part of a physical unit
e.g. qubit, to another part is a crucial ingredient for many quantum
information processing protocols \cite{Bouwmeester_book}. In fact,
the ability to transmit ``flying qubits'' is known as one of the
DiVincenzo's desiderata \cite{DiVincenzo_Fort} for quantum
communication. For long distance communication, it is natural to
rely on optical means \cite{Cirac97}. However, for systems involved
in a sufficiently small scale where qubits can interact either
directly or indirectly through other qubits, it seems to be more
natural to exploit the interactions between them directly. In these
scenarios, a typical assumption usually made is that the
interactions between pairs of qubits can be arbitrarily switched on
and off \cite{{Kane98},{DiVincenzo00}}. In this way, a quantum state
can be transmitted by a series of swap (SOS) operations. However,
even in the absence of decoherence, the intrinsic problem of this
protocol is that each swap would necessarily introduce some errors
due to dynamical controls, the fidelity of the transferred states
would therefore decay exponentially as the number of swaps
increases. It is therefore of great practical interest to study
schemes of perfect state transfer by \emph{free} Hamiltonian
evolution which is primarily designed to minimize dynamical controls
as much as possible.

This work is also motivated by the question of the ultimate speed
limit of solid-state quantum computers. The Margolus-Levitin theorem
\cite{Margolus98} suggests that the shortest time for a quantum
state to evolve into an \emph{orthogonal} state is limited by the
initial energy. With this result, one can quantify the speed limit
for elementary operations involving one or two qubits
\cite{Lloyd00}. However, this theorem does not give an appropriate
bound on the speed limit for the initial state to evolve into a
\emph{particular} state, for example the one dimensional quantum
state transfer (QST) problem [cf. Eq. (\ref{U})]. The aim of this
paper is to establish the speed limit for the QST problem. Following
the scheme described in Ref. \cite{Yung05}, one can generalize the
study to more general quantum operations. In contrast with the
Margolus-Levitin theorem, the quantum speed limit for QST will be
characterized directly by the couplings in the Hamiltonian, instead
of the initial energy. As we shall see, this makes it not only
experimentally more accessible, not also physically more reasonable.

Imperfect QST \cite{Bose03,Yung03} over a uniform spin chain was
studied numerically, showing that perfect state transfer is
impossible for long chains. However, it has been suggested
\cite{Christandl04} that perfect QST, without dynamical controls, is
possible if we allow the couplings between the spins to be
non-uniform. Moreover, it has also been pointed out \cite{Yung04}
that there are many possible solutions for constructing spin chains
which allow perfect state transfer, and the searching for the
solutions is in general an inverse eigenvalue problem. Recently,
there have been many quantum information processing proposals
\cite{Yung05,Albanese04,Cubitt05,Shi05,Yang05,Kay05,Kay05a,Clark05,Karbach05,Chiara05}
related to the concept of spin chain engineering. To boost the
performance of the spin chains, or to evaluate the ultimate speed
limit of these proposals, it is highly desirable to know how to
engineer spin chains with optimal performance.

In the following, we shall show that the optimal solution for
perfect state transfer is exactly the one proposed in
Ref.\cite{Christandl04,footnote1}. Amusingly, this optimal set of
couplings is the same as that describing spin precession in a static
magnetic field. In discussing the one dimensional QST problem, one
usually considers the Hamiltonian for the engineered spin-1/2 chains
to be of the following form \cite{footnote2}:
\begin{equation}\label{H}
H = \sum\limits_{j = 1}^{N - 1} {\frac{{\omega _j }}{2}(\sigma _j^x
\sigma _{j + 1}^x  + \sigma _j^y \sigma _{j + 1}^y )}  +
\sum\limits_{j = 1}^{N } {\frac{{\lambda _j }}{2}(\sigma _j^z  + 1)}
\, ,
\end{equation}
where $\sigma _j$'s are the standard Pauli matrices for the site
$j$, and $N$ is the total number of sites in the spin chain. Both
$\omega _j$ and ${\lambda _j }$ are real constants to be determined.
We shall adopt the convention that $\left| {0_j } \right\rangle$
($\left| {1_j } \right\rangle$) represents spin-down (spin-up)
state.

In the quantum state transfer protocol, one site is encoded with a
quantum state $\alpha \left| 1 \right\rangle + \beta \left| 0
\right\rangle$ representing the information of one qubit, while the
rest are initialized to be spin-down. Define $U\left( \tau  \right)
= e^{ - iH\tau }$, where $\tau$ is the evolution time and $\hbar
=1$. Since $U\left( \tau \right)\left| {000...00} \right\rangle =
\left| {000...00} \right\rangle$, the state is said to be perfectly
transferred whenever
\begin{equation}\label{U}
U\left( \tau \right)\left| {100...00} \right\rangle  = e^{i\phi }
\left| {000...01} \right\rangle \quad ,
\end{equation}
where $\phi$ is some known phase. Although the transferred state in
general differs from the input state by a relative phase factor,
i.e. $\alpha e^{i\phi } \left| 1 \right\rangle + \beta \left| 0
\right\rangle$, the phase factor could be corrected by some local
operations and is thus usually ignored. To investigate further, it
is clear that we can simply focus on the single-particle subspace
$\left\{ {\left| {\bm 1} \right\rangle \equiv \left| {100...0}
\right\rangle ,\left| {\bm 2} \right\rangle \equiv \left| {010...0}
\right\rangle ,...} \right\}$, where ${\left| {\bm x} \right\rangle
}$, $x=1,2,3,..,N$ refers to a state with a single excitation at
site $x$. In this subspace, the Hamiltonian in Eq. (\ref{H}) is
real, symmetrical and tridiagonal:
\begin{equation}
H_S = \left( {\begin{array}{*{20}c}\label{HS}
   {\lambda _1 } & {\omega _1 } & 0 &  \cdots  & 0  \\
   {\omega _1 } & {\lambda _2 } & {\omega _2 } &  \cdots  & 0  \\
   0 & {\omega _2 } & {\lambda _3 } &  \cdots  & 0  \\
    \vdots  &  \vdots  &  \vdots  &  \ddots  & {\omega _{N - 1} }  \\
   0 & 0 & 0 & {\omega _{N - 1} } & {\lambda _N }  \\
\end{array}} \right) \quad .
\end{equation}
The eigenvalues, which must be non-degenerate \cite{GladwellBOOK},
and the corresponding eigenvectors are denoted by $E_k$ and $\left|
{\bm e}_k \right\rangle$ respectively. Let us consider not only
transferring states from one end to the other but more generally the
transition amplitude between the state $\left| \bm x \right\rangle$
and its mirror-inverted state $\left| \bar {\bm x} \right\rangle$,
with $\bar x{\equiv} N{-}x {+} 1$,
\begin{equation}\label{map}
\left\langle {\bar {\bm x}} \right| U ( \tau )\left| \bm x
\right\rangle = \sum\limits_{k = 1}^N {} \langle {\bar {\bm x}}
|{{\bm e}_k }\rangle \langle {{\bm e}_k } | {\bm x} \rangle e^{ -
iE_k \tau } \quad .
\end{equation}
Provided that the Hamiltonian in Eq.~(\ref{H}) or Eq.~(\ref{HS}) is
mirror symmetrical, i.e. $\lambda _j  = \lambda _{\bar j}$ and
$\omega _j = \omega _{N - j}$, one can show \cite{Yung04} that
$\langle {\bar {\bm x}} | {\bm e}_k  \rangle  = \left( { - 1}
\right)^k \langle {\bm x} |{{\bm e}_k } \rangle$, and hence $\left|
{\left\langle {\bar x} \right|U\left( \tau  \right)\left| x
\right\rangle } \right| = 1$ if
\begin{equation}\label{condition}
e^{ - iE_k \tau }  = \left( { - 1} \right)^k e^{i\phi } \quad .
\end{equation}
The phase factor $e^{i\phi }$ here is exactly the same as that
appeared in Eq. (\ref{U}). It has been shown \cite{Yung04} that the
condition in Eq. (\ref{condition}) is not only sufficient but also
necessary for perfect state transfer in mirror symmetric chains.
Given an eigenvalue spectrum $\{E_k\}$ satisfying Eq.
(\ref{condition}), the corresponding coupling constants
$\{\omega_j,\lambda_j\}$ can be uniquely determined. The task of
solving for the solutions of $\{\omega_j,\lambda_j\}$ is therefore
an inverse eigenvalue problem \cite{Boley87,{GladwellBOOK}}, which
generally has to be solved numerically. A spin chain is said to be
engineered if its couplings are found in this way.

Perhaps the simplest solution to the Eq. (\ref{condition}) is the
linear spectrum \cite{Christandl04}. By mapping the states $\left|
\bm x \right\rangle$ to be the eigenstates of the $J_z$ angular
momentum operator, this solution also describes the precession of a
spin $J{=}(N{-}1)/2$ under a constant magnetic field pointing along
the $x$-direction. Explicitly, the matrix elements of $H_S$ can be
chosen as
\begin{equation}\label{optimal}
\lambda _j  = 0\quad \mathrm{and} \quad \omega _j  =
\frac{1}{2}\sqrt {j\left( {N - j} \right)} \quad .
\end{equation}
The eigenvalue spectrum is linear in the sense that the energy
spacing is uniform, $ E_k {=} -({N {-} 1}){/}2{+}k {-} 1$,
$k=1,2,3,...,N$. From Eq. (\ref{map}) and (\ref{condition}), perfect
state transfer can be achieved for a time period of $\tau=\pi$.

Apparently, there are unlimited number of eigenvalue spectra
satisfying the condition in Eq. (\ref{condition}). A number of
particular solutions have also been found
\cite{Albanese04,Yung04,Shi05,Karbach05} recently. To compare
different sets of inter-qubit coupling constants $\left\{ {\omega _j
} \right\}$ generated from different energy eigenvalue spectra
$\{E_k\}$, the efficiency $\eta$ can be quantified by the evolution
time $\tau$ for completing the task of perfect state transfer,
subject to the constraint that the maximum value of inter-spin
coupling $\omega _{\max }  \equiv \max \left\{ {\omega _j }
\right\}$ being normalized. Alternatively, one may fix the evolution
time and compare the maximum coupling strength for different chains.
Both cases can be properly captured by defining the efficiency as
\begin{equation}\label{eta}
\eta  = \frac{{\omega _{\max } \tau }}{{\tilde \omega _{\max }
\tilde \tau }} \quad ,
\end{equation}
where ${\tilde \omega _{\max } }$ and ${\tilde \tau }$ are
respectively the maximum coupling and the evolution time of a
reference spin chain, which we shall choose to be the one described
in Eq. (\ref{optimal}), i.e. $\tilde \tau  = \pi$ and
\begin{equation}\label{omega_max}
\tilde \omega _{\max }  = \left\{ {\begin{array}{*{20}c}
      {\frac{1}{4}\sqrt {N^2  - 1} } & {{\rm{for \enspace odd \enspace }}N},  \\
      {\frac{1}{4}N} & {{\rm{for \enspace even \enspace }}N}.  \\
\end{array}} \right.
\end{equation}
With these definitions, our goal is to show that $\eta  \geq 1$ for
all engineered spin chains. This implies that the set of couplings
in Eq. (\ref{optimal}) is optimal and the quantum speed limit for
one dimensional QST is established (i.e. given the same maximum
couplings $\omega _{\max }  = \tilde \omega _{\max }$, $\tau  = \eta
\tilde \tau  \ge \tilde \tau $). A comparison for different schemes
of perfect state transfer is given in Table \ref{table1}.

\begin{table}[t]
\caption{\label{table1} Comparing the efficiency $\eta$ (defined in
Eq. (\ref{eta})) for various possible solutions to the problem of
perfect state transfer through a spin chain of $N$ spins. For
convenience, we have scaled the maximum tunneling matrix element
$\omega_{\max}$ such that the minimal evolution time $\tau=\pi$ is
fixed.}
\begin{ruledtabular}
\begin{tabular}{lll}
Type of Spectrum & $\omega_{\max}$ & Efficiency $\eta$\\
\hline

Linear\footnote{See Ref.\cite{Christandl04}. $\tilde \omega _{\max
}$ is defined in Eq.(\ref{omega_max}) and $\eta=1$ by definition.} &
$\tilde \omega _{\max
}$  & 1 \\

Quadratic\footnote{See Ref.\cite{Albanese04}. For $N^2{\gg}1$,
$q\left( { \alpha {+} 1} \right) {\ge} 1/2$ as required by
$E_1{-}E_0 {\ge} 1$, where $q$ is a positive integer and $\alpha$ is
a rational number.} & $ q \left( {2 \alpha {+} N} \right) \tilde
\omega _{\max } $ & $ q
\left( {2 \alpha {+} N} \right)$ \\

Gapped linear I\footnote{See Ref.\cite{Shi05}. For $N^2\gg1$.} &
$\left( {1 {+} 4q/N} \right) \tilde \omega _{\max } $ & $\left( {1 {+} 4q/N} \right)$\\

Gapped linear II\footnote{See Ref.\cite{Yung04}. For $N=4$ only.} & $q$ & $q$\\

Cosine\footnote{See Ref.\cite{Karbach05}. For $N=31$ (no analytic solution given).}  & 108.5 & 14.0\\

\end{tabular}
\end{ruledtabular}
\end{table}

We start with the observation pointed out in Ref. \cite{Yung05}.
Suppose the eigenvalues are ordered as $E_1  > E_2
>\cdots >E_N$. Define $\Delta _k  \equiv E_k  - E_{k + 1}$ and the range of
the spectrum as $\Delta_E \equiv E_1  - E_N  = \sum\nolimits_{k =
1}^{N - 1} {\Delta _k }$ . From Eq. (\ref{condition}), the evolution
time is limited by the minimum energy interval $\Delta _{\min }
\equiv \min \left\{ {\Delta _k } \right\}$, $\tau \ge \pi / \Delta
_{\min }$, while $\tilde \tau  = \pi /\tilde \Delta _{\min }$. From
Eq. (\ref{eta}), we have
\begin{equation}
\eta  \ge \frac{{\omega _{\max } \tilde \Delta _{\min } }}{{\tilde
\omega _{\max } \Delta _{\min } }} \quad .
\end{equation}
Suppose we scale the eigenvalue spectra such that $\tilde \Delta
_{\min }  = \Delta _{\min }$, then $\eta  \ge 1$ if $\omega _{\max }
\ge \tilde \omega _{\max }$. We shall show that it is indeed true.
%

In the single-spin subspace, applying the Hellmann-Feynman theorem
to the Hamiltonian in Eq. (\ref{H}), we have $\partial E_k /\partial
\lambda _j= \frac{1}{2}\left\langle {{\bm e}_k } \right|\left(
{\sigma _j^z + 1} \right)\left| {{\bm e}_k } \right\rangle$ and
$\partial E_k /\partial \omega _j = 2 \langle {{\bm e}_k } | {\bm j}
\rangle \langle {\bm j} {+} {\bm 1} | {{\bm e}_k } \rangle $. It is
clear that $\partial E_k /\partial \lambda _j  \ge 0$. On the other
hand, it is known \cite{GladwellBOOK} that the eigenvector $| {{\bm
e}_k } \rangle$ has exactly $k-1$ sign changes in the position basis
$\left\{ {\left| {\bm j} \right\rangle } \right\}$. In other words,
$\partial E_1 /\partial \omega _j  \ge 0 $ and $\partial E_N
/\partial \omega _j  \le 0$. Suppose we replace all $\omega _j \to
\omega _{\max }$ and $\lambda _j  \to \lambda _{\max } ,\lambda
_{\min }$ respectively, and invoke the solution of the eigenvalues
for a uniformly coupled chain: $\lambda + 2 \omega \cos (
{\frac{{k\pi }}{{N + 1}}})$, we have
\begin{equation}\label{inequality}
\Delta _E  \le \Delta _\lambda   + 4 \omega _{\max } \quad ,
\end{equation}
where $\Delta _\lambda   \equiv \lambda _{\max }  - \lambda _{\min
}$ is the range of the spatial variations of the local potentials.
Suppose, in the absence of, or with uniform, local potentials, we
set $\Delta _\lambda   = 0$. Under the constraint that the evolution
time to be the same, i.e. $\Delta _{\min } \ge \tilde \Delta _{\min
} \equiv 1$, the range of any nonlinear spectrum must be at least
greater then that of the linear spectrum by one, $\Delta _E \ge
\tilde \Delta _E+1=N $, as the ratio between any two $\Delta _k $'s
must be a rational number \cite{Christandl04}. One can then show
from Eq. (\ref{omega_max}) and (\ref{inequality}) that $\omega
_{\max } \ge \tilde \omega _{\max }$. Of course, this argument is
based on the assumption that the terms $\lambda _j$ play no role in
the eigenvalue spectrum except a constant shift. The question is:
could the speed limit be boosted, if spatially varying local
potentials are allowed? The aim of the following paragraphs is to
exclude such possibility, and at the same time provide a more
rigorous proof.

Exploiting the symmetry of our problem, we can divide the $H_S$ in
Eq. (\ref{HS}) into two different subspaces \cite{Yung03,Yung04}.
For even $N$, they are simply $\frac{1}{{\sqrt 2 }}\left\{ {\left|
{\bm j} \right\rangle  + \left| \bm {\bar j } \right\rangle }
\right\}$ and  $\frac{1}{{\sqrt 2 }}\left\{ {\left| {\bm j}
\right\rangle  - \left| \bm {\bar j } \right\rangle } \right\}$,
$j=1,2,..,N/2$. The $N \times N$ Hamiltonian reduces into two $
{\textstyle{N \over 2}} \times {\textstyle{N \over 2}}$ ones:
\begin{equation}
\left( {\begin{array}{*{20}c}
   {\lambda _1 } & {\omega _1 } & 0 &  \cdots  & 0  \\
   {\omega _1 } & {\lambda _2 } & {\omega _2 } &  \cdots  & 0  \\
   0 & {\omega _2 } & {\lambda _3 } &  \cdots  & 0  \\
    \vdots  &  \vdots  &  \vdots  &  \ddots  & {\omega _{N/2 - 1} }  \\
   0 & 0 & 0 & {\omega _{N{/}2 - 1} } & {\lambda _{N/2}  \pm \omega _{N/2} }  \\
\end{array}} \right) \quad ,
\end{equation}
These two matrices are almost identical except one diagonal matrix
element ${\lambda _{N/2}  \pm \omega _{N/2} }$. Recall that $\langle
{\bar {\bm x}} | {\bm e}_k  \rangle = \left( { - 1} \right)^k
\langle {\bm x} |{{\bm e}_k } \rangle$, the eigenvectors of $H_S$
are automatically grouped into these two subspaces. In other words,
the eigenvalue spectra of these two matrices are respectively $\nu
_k = \left\{ {E_1 ,E_3 ,E_5 ,...,E_{N-1}} \right\}$ and $\mu _k  =
\left\{ {E_2 ,E_4 ,E_6 ,...,E_N } \right\}$. Consequently, the trace
difference between the two matrices gives $\omega _{N/2}  = \left(
{1/2} \right)\sum\nolimits_k {\left( {\nu _k  - \mu _k } \right)}  =
\left( {1/2} \right)\sum\nolimits_k {\Delta _{2k-1} }$, which is
minimized when all $\Delta _{2k-1}  = \tilde \Delta _{\min }$, which
corresponds to the case $\omega _{N/2}  = \tilde \omega _{\max }$.
Hence we can conclude that $\omega _{\max }  \ge \omega _{N/2} \ge
\tilde \omega _{\max }$ for even $N$.

For odd $N {\ge} 5$ (the case $N{=}3$ can be analytically solved
separately), it becomes slightly more complicated. The subspaces are
spanned by $\frac{1}{{\sqrt 2 }}\left\{ {\left| \bm j \right\rangle
- \left| \bm {\bar j} \right\rangle } \right\}$, and
$\frac{1}{{\sqrt 2 }}\left\{ {\left| \bm j \right\rangle  + \left|
\bm {\bar j} \right\rangle } \right\}$ together with $\left| \bm m
\right\rangle$, where $j = 1,2,3,...,m{-}1$ and $m \equiv \left( {N
{+} 1} \right)/2$. The matrix formed by the latter is
\begin{equation}\label{Matrix}
\left( {\begin{array}{*{20}c}
   {\lambda _1 } & {\omega _1 } & 0 & 0 & 0  \\
   {\omega _1 } &  \ddots  &  \vdots  &  \vdots  &  \vdots   \\
   0 &  \cdots  & {\lambda _{m - 2} } & {\omega _{m - 1} } & 0  \\
   0 &  \cdots  & {\omega _{m - 1} } & {\lambda _{m - 1} } & {\sqrt 2 \omega _m }  \\
   0 &  \cdots  & 0 & {\sqrt 2 \omega _m } & {\lambda _m }  \\
\end{array}} \right) \quad ,
\end{equation}
and the counterpart by the former can be obtained from the above
matrix by removing the last row and the last column. Similar to the
case of even $N$, the eigenvalue spectra are $\nu _k  = \left\{ {E_1
,E_3 ,...,E_N } \right\}$ and $\mu _k  = \left\{ {E_2 ,E_4 ,...,E_{N
- 1} } \right\}$ respectively. Our goal is still to show $\omega _m
\ge \tilde \omega _{\max }$ when $\Delta _{\min }  \ge \tilde \Delta
_{\min }$. To proceed, one can diagonalize the upper part of the
matrix in Eq. (\ref{Matrix}), i.e. the part from $\lambda _1$ to
$\lambda _{m-1}$ , and the resulting matrix is of the bordered
diagonal form (arrowhead matrix) \cite{Boley87}
\begin{equation}
\left( {\begin{array}{*{20}c}
   {\lambda _1 } & 0 &  \cdots  & 0 & {b_1 }  \\
   0 & {\lambda _2 } &  \cdots  & 0 & {b_2 }  \\
    \vdots  &  \vdots  &  \ddots  & 0 &  \vdots   \\
   0 & 0 & 0 & {\lambda _{m - 1} } & {b_{m - 1} }  \\
   {b_1 } & {b_2 } &  \cdots  & {b_{m - 1} } & a  \\
\end{array}} \right) \quad ,
\end{equation}
where $a = \lambda_m = \sum\nolimits_{k = 1}^m {\nu _k }  -
\sum\nolimits_{k = 1}^{m - 1} {\mu _k }$ is the trace difference,
and $b_k$ are the off-diagonal matrix elements satisfying the
condition $2 \omega _m^2 = \sum\nolimits_{k = 1}^{m - 1} {b_k^2 }$,
as a result of of the diagonalization.

The characteristic polynomial $P\left( E \right) = \prod\nolimits_{k
= 1}^m {\left( {E - \nu _k } \right)}$ of the matrix above can be
written as
\begin{equation}\label{PQ}
P\left( E \right) = \left( {E - a} \right)Q\left( E \right) -
\sum\limits_{k=1}^{m-1} {\frac{{b_k^2 }}{{\left( {E - \mu _k }
\right)}}} Q\left( E \right) \quad ,
\end{equation}
where $Q\left( E \right) = \prod\nolimits_{k = 1}^{m - 1} {\left( {E
- \mu _k } \right)}$. By setting $E=\mu _k$, we have
\begin{equation}\label{bk}
b_k^2  =  - \frac{{\prod\nolimits_{j = 1}^m {\left( {\mu _k  - \nu
_j } \right)} }}{{\prod\nolimits_{j \ne k}^{m - 1} {\left( {\mu _k -
\mu _j } \right)} }} \ge 0 \quad ,
\end{equation}
where the negative sign here is because of the interlacing property
$\nu _k  > \mu _k$. One can solve for $\omega _m$ recursively using
Eq. (\ref{PQ}) and (\ref{bk}). We found a recursive relation $\omega
_m^2 = {\bar \omega} _{m - 1}^2  + \frac{1}{2}\Delta _{N - 1} \left(
{\sum\nolimits_{k = 1}^{(N - 1)/2} {\Delta _{2k - 1} } } \right)$,
where $\bar \omega _{m - 1}$ is the counterpart of $\omega_m$ when
we solve the same inverse eigenvalue problem without the last two
eigenvalues $E_{N-1}$ and $E_N$. Since $\omega _m^2$ is a positive
sum of the products $\Delta _j \Delta _k$, it is minimized when all
$\Delta _k = \tilde \Delta _{\min }$, and hence $\omega _{\max } \ge
\omega _m \ge \tilde \omega _{\max }$. This completes our proof for
$\eta \ge 1$ for all possible engineered spin chains.

In the following, we shall analyze the implications of the speed
limit. It is natural to ask, apart from the advantage of not
requiring dynamical controls, if the schemes of engineered chains
can be fundamentally faster than the series of swaps (SOS) protocol
mentioned at the beginning. The answer is positive. Suppose we now
consider the ideal case where each swap operation can be achieved
with the \emph{same} maximum possible coupling $\tilde
\omega_{\max}$ in Eq. (\ref{omega_max}), the evolution time for each
swap is $\tau _0 = \pi /2\tilde \omega _{\max }$. Since it will
totally take $N-1$ steps to transfer a state through a chain of $N$
spins, we can generalize our definition of the efficiency for this
case as: $\eta _{sos} = \left( {N - 1} \right)/2\tilde \omega _{\max
}$, which is an increasing function of $N$, and one can easily show
that $1 \le \eta _{sos}  < 2$ for $N\ge2$. Therefore, the engineered
spin chains can have roughly a factor of two gain in speed for
sufficiently long chains.

According to the Margolus-Levitin theorem \cite{Margolus98}, the
minimum time for a quantum state  $|{\psi_i }\rangle$ to evolve into
an orthogonal state $| {\psi_f }\rangle$, where $\langle {\psi _f }|
{\psi _i }\rangle {=} 0$, is limited by the initial energy: $ \max
\left( {\frac{\pi }{{2( {E {-} E_0 })}},\frac{\pi }{{2\Delta E}}}
\right)$, where $E {=} \langle {\psi _i } |H| {\psi _i }\rangle$ is
the initial energy, $E_0$ is the ground state energy and $\Delta E
{=} \sqrt {\langle {\psi _i } |( {H - E} )^2 | {\psi _i }\rangle }$
is the energy uncertainty of the initial state. The computational
power of physical systems as quantum computers was analyzed
\cite{Lloyd00} along this line of thought. Here we shall argue that,
at least for solid state systems where the interactions are local,
it is more realistic to characterize the computational power by the
couplings in the Hamiltonian than the initial energy. We shall again
focus on the 1D QST problem [cf. Eq.(\ref{U})] as an example, as the
generalization to more complicated quantum operations is
straightforward through the scheme proposed in Ref. \cite{Yung05}.
For the linear spectrum, we have $E {-} E_0 {=} N/2$ for large $N$.
Given the same amount of initial energy $N/2$, the minimal evolution
time for the SOS protocol is $\tau = \pi /N$ (as $E {-} E_0  =
\Delta E$). For a linear chain of $N$ spins, the SOS protocol seems
to be as efficient as the engineered chains. However, as we have
just analyzed, for each swap the tunneling strength would be $N/2$,
which is twice of the maximum possible coupling $\tilde
\omega_{\max}$ assumed before!

In conclusion, we have generalized the study of quantum speed limit
to the problem of quantum state transfer in one dimension. We have
argued that for solid state systems it is more realistic to
characterize the computation power explicitly by the coupling terms
in the Hamiltonian than the initial energy. This work also implies
that quantum algorithms, when implemented in solid state devices,
can be optimized, if we employ the concept of spin chain engineering
\cite{Yung05}. More practically, systematic errors are expected to
be significantly minimized with the reduction of dynamical controls.
We lastly comment that such speed limit, if not confined with
nearest-neighbor interactions in one dimension, could be overcome by
increasing either the network complexity \cite{Yung03} or the
Hamiltonian complexity \cite{Giovannetti}. Details to be discussed
elsewhere.



\begin{acknowledgments}
MHY acknowledges the support of the NSF grant DMR-03-50842  and the
Croucher Foundation and thanks S.~C.~Benjamin, S.~Bose, Alec Maassen
van den Brink and L.~Maccone for valuable discussions.
\end{acknowledgments}




\begin{thebibliography}{99}


\bibitem{Bouwmeester_book}
D. Bouwmeester , A. Ekert, and A. Zeilinger, \emph{The Physics of
Quantum Information} (Springer-Verlag, Berlin, 2000).

\bibitem{DiVincenzo_Fort}
D.~P.~DiVincenzo, Fortschr. Phys. \textbf{48}, 9 (2000).

\bibitem{Cirac97}
J.~I.~Cirac, P.~Zoller, H.~J.~Kimble, and H.~Mabuchi, Phys. Rev.
Lett. \textbf{78}, 3221 (1997).

\bibitem{Kane98}
B.~E.~Kane, Nature (London) \textbf{393}, 133 (1998).

\bibitem{DiVincenzo00}
In many systems, even though the two-body interactions are
intrinsically not switchable, their effects are usually made to be
annulled by some refocusing techniques.

\bibitem{Margolus98}
N.~Margolus and L.~B.~Levitin, Physica D \textbf{120}, 188 (1998);
P.~Kosi\'{n}ski and M.~Zych, Phys. Rev. A \textbf{73}, 024303
(2006).

\bibitem{Lloyd00}
S.~Lloyd, Nature (London) \textbf{406}, 1047 (2000).

\bibitem{Bose03}
S. Bose, Phys. Rev. Lett. \textbf{91}, 207901 (2003).

\bibitem{Yung03}
M.-H.~Yung, D. W. Leung, and S. Bose, Quantum Inf. Commun.
\textbf{4}, 174 (2003).

\bibitem{Yung04}
M.-H.~Yung and S. Bose, Phys. Rev. A \textbf{71}, 032310 (2005).

\bibitem{Yung05}
M.-H.~Yung, S.~C.~Benjamin and S.~Bose, e-print quant-ph/0508165 (To
appear in Physical Review Letter).

\bibitem{Christandl04}
M.~Christandl, N.~Datta, A.~Ekert and A.~J.~Landahl, Phys. Rev.
Lett. \textbf{92}, 187902 (2004); M.~Christandl, N.~Datta,
T.~C.~Dorlas, A.~Ekert, A.~Kay, and A.~J.~Landahl, Phys. Rev. A
\textbf{71}, 032312 (2005).

\bibitem{Albanese04}
C. Albanese, M. Christandl, N. Datta and A. Ekert, Phys. Rev. Lett.
\textbf{93}, 230502 (2004).

\bibitem{Cubitt05}
T.~S.~Cubitt, F.~Verstraete, and J.~I.~Cirac, Phys. Rev. A
\textbf{71}, 052308 (2005).

\bibitem{Shi05}
T.~Shi, Y.~Li, Z.~Song, and C.~P.~Sun, Phys. Rev. A \textbf{71},
032309 (2005).

\bibitem{Yang05}
S.~Yang, Z.~Song and C.~P.~Sun, e-print quant-ph/0510109.

\bibitem{Kay05}
A.~Kay and M.~Ericsson, New J. Phys. \textbf{7}, 143 (2005).

\bibitem{Kay05a}
A.~Kay, Phys. Rev. A \textbf{73}, 032306 (2006).

\bibitem{Clark05}
S. R. Clark, C. Moura Alves and D. Jaksch, New J. Phys. \textbf{7},
124 (2005).

\bibitem{Karbach05}
P.~Karbach and J.~Stolze, Phys. Rev. A \textbf{72}, 030301(R)
(2005).

\bibitem{Chiara05}
G.~De~Chiara, D.~Rossini, S.~Montangero and R.~Fazio,  Phys. Rev. A
\textbf{72}, 012323 (2005).

\bibitem{footnote1}
As pointed out in Ref.\cite{Kay05}, the same set of coupling was
first proposed by Cook and Shore in 1979 [R.~J.~Cook and
B.~W.~Shore, Phys.~Rev.~A \textbf{20}, 539 (1979)] in the context of
coherent population transfer between different atomic levels.

\bibitem{footnote2}
Our result on the quantum speed limit is not necessarily confined by
the exact form of this Hamiltonian. In fact, it is generally
applicable to many hopping processes of spins or particles, as long
as the effective Hamiltonian is of the form described in Eq.
(\ref{HS}).

\bibitem{Boley87}
D. Boley and G. H. Golub, Inverse Problems \textbf{3}, 595 (1987).

\bibitem{GladwellBOOK}
G.~M.~L.~Gladwell, \emph{Inverse problems in vibration} (Kluwer
Academic, Boston, 1986).

\bibitem{Giovannetti}
V.~Giovannetti, S.~Lloyd, L.~Maccone, Europhys. Lett. \textbf{62},
615 (2003).

\end{thebibliography}
\end{document}